\documentclass[12pt]{iopart}
\usepackage{graphicx}
\usepackage{float}

\begin{document}

\title[Stochastic analysis of ocean wave states]{Stochastic analysis of ocean wave states with and without rogue waves}

\author{A. Hadjihosseini$^{1}$, J. Peinke$^{1}$, N. P. Hoffmann$^{2,3}$}

\address{$^{1}$ Universit\"at Oldenburg, 26111 Oldenburg, Germany\\
$^{2}$ Hamburg University of Technology, 21073 Hamburg, Germany\\
$^{3}$ Imperial College, London SW7 2AZ, United Kingdom\\}
\ead{\mailto{ali.hadjihosseini@uni-oldenburg.de}}
\begin{abstract}
This work presents an analysis of ocean wave data including rogue waves. A stochastic approach based on the theory of Markov processes is applied. With this analysis we achieve a characterization of the scale dependent complexity of ocean waves by means of a Fokker-Planck equation, providing stochastic information of multi-scale processes. In particular we show evidence of Markov properties for increment processes, which means that a three point closure for the complexity of the wave structures seems to be valid. Furthermore we estimate the parameters of the Fokker-Planck equation by parameter-free data analysis. The resulting Fokker-Planck equations are verified by numerical reconstruction. This work presents a new approach where the coherent structure of rogue waves seems to be integrated into the fundamental statistics of complex wave states.

\end{abstract}

\submitto{\NJP}
\maketitle

\section{Introduction}
Rogue or freak waves are exceptionally large waves which at present are subject of intense studies in a number of different fields \cite{Chabchoub2011,Solli2007,Akhmediev2010,Moslem2011,BENCHRIET2013,Veldes2013,Dudley2010,zakharov2010}. In the ocean waves are today called rogue waves if their amplitude exceeds twice (some also state 2.2 or 2.5 times) that of the significant wave height which is defined as the average height of the one-third largest waves. Most remarkably, rogue waves are often considered to appear seemingly from nowhere and to disappear without a trace\cite{Akhmediev2010,Akhmediev2009}, and thus pose substantial threat to ships and offshore structures. Installations extracting energy from or on the ocean, like offshore wind farms, wave energy conversion devices or ocean tidal and ocean current turbines seem especially concerned, since they have to be able to withstand rogue waves and to maintain the continuity of the energy supply to remain economically viable.

At present it is under intense and controversial discussion if the physics and the mathematical description of rogue wave phenomena in the different disciplines have a generic or common ground. Especially recent work based on the Nonlinear Schr\"odinger Equation (NLSE), describing the weakly nonlinear dynamics of narrow-banded wave states, suggests that nonlinear breather states \cite{Bludov2009} might play a crucial role and might form the deterministic backbone of nonlinear focussing of wave energy in a background wave state. Opposed to this hypothesis, which is fundamentally nonlinear in nature, alternative linear mechanisms are also under study, and the picture is far from complete. Also in statistical physics an interest in rogue waves has been arising recently in the context of the study of extreme events. Studies on extreme events are often focussed on predicting probability density functions (pdf) of event characteristics, and links to stochastic theory \cite{Fedele2007} and nonlinear random wave theory \cite{Onorato2002,Onorato2001,Fedele2006,Onorato2009} do exist.

In the present study we make an attempt to characterize ocean wave states by applying and extending a technique to identify stochastic differential equations from measured data. One of the difficulties in studies on rogue waves within a natural sea state of the ocean is caused by their rare occurrence in field measurements. In consequence data based information about rogue waves is comparatively poor and incomplete, and definitely more analysis of real ocean wave measurements is highly desirable. Today there is only a limited number of publications presenting analysis of extreme waves recorded in the North Sea \cite{Soares2003,Haver1998,Sand1990}, the Japan Sea \cite{Mori2002a}, and the Black Sea \cite{Divinsky2004}. The number of laboratory experiments in wave tanks to study wave structures, wave formation, and to test different hypotheses on wave turbulence and rogue wave formation, is also relatively small \cite{Chabchoub2011}. Due to the often highly idealized setup of wave experiments, in the present study we will apply our techniques to ocean wave data only.

But also field measurements have to be considered with great care. There are several in-situ instrumentation techniques, based on different measurement principles, that can track the surface elevation of the water, including laser and radar altimeters, buoys, and subsurface instruments such as pressure gauges and acoustic devices. The performance of several different measurement systems was e.g. compared in the WACSIS experiment at the Dutch coast and at the Tern platform in the North Sea \cite{Forristall2004}. Data errors and inaccuracies are particularly serious when looking for exceptionally large waves. Spikes in the data due to measurement error may be mistaken for rogue waves and in addition many recording systems (e.g. the traditional wave buoys) employ mechanical and electronic filters that need to be compensated for when an accurate tracking of the actual surface elevation is required. In some exceptional cases, measurements are, however, supported by independent and observable physical effects. This was e.g. the case for the Draupner incident, in which damage was reported to equipment on a temporary deck below the main deck of the platform. In the present study we will apply the technique to a wave data set, measured in the sea of Japan, that we consider very trustworthy. The data contains also some waves that can be classified as rogue waves, and so the analysis presented is hopefully also shedding some further light on the question of how rogue waves are embedded into an irregular natural background sea state. 

The approach that we apply has originally been introduced by Friedrich $\&$ Peinke \cite{Friedrich1997} and since then has been successfully applied in a large variety of fields, such as turbulence \cite{Friedrich1997,Renner2001b}, economics\cite{Renner2001a,Nawroth2010,Ghasemi2007}, biology\cite{Friedrich2000,Ghasemi2006,Atyabi2006}, and many more, see \cite{Friedrich2011}. It is based on identifying and exploiting Markov properties for the evolution of probability density functions, and starts out with measured data to identify coefficients of fundamental stochastic differential equations, like e.g. Fokker-Planck equations. The properties of the resulting equations in turn characterize the complex data. An important new aspect is that stochastic processes not only evolving in time but also in relative scale variable can be grasped by this procedure. With scale dependent processes fractal and multi-fractal structures and even more generally joint n-point statistics can be reduced by the Markow properties to particular three point statistics,  cf. \cite{Stresing2010,Friedrich2011}.

The paper is structured as follows. First, some fundamental statistical properties and spectral properties of the data set under study are presented. Then the present approach is described. Subsequently it is shown how a Markovian process for the pdf of the surface elevation can be derived from the data, how Kramers-Moyal coefficients can be estimated, and how a Fokker-Planck equation governing conditional probability densities can be derived that contains a dependency on the time-scale of the process, by which the analysis allows insight into the temporal multi-scale nature of irregular ocean waves, i.e. a natural seaway.

\section{Measurement data and probability distribution functions}

\begin{figure}[H]
  \begin{center}
    \includegraphics[width=.95\textwidth , height=6 cm]{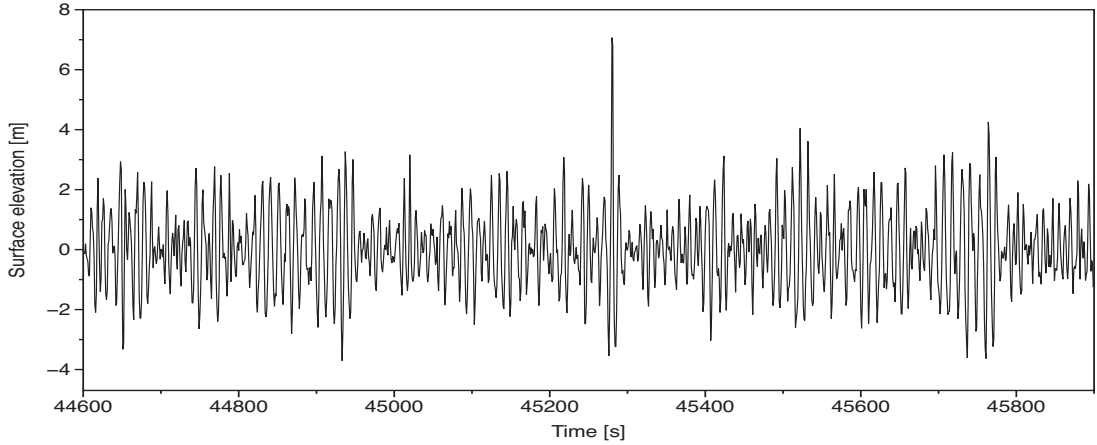}
  \end{center}
 \caption{Part of the surface elevation recordings in the Japan sea. Some waves fulfilling the rogue wave definition are included.}
 \label{Yura}
\end{figure}

The wave measurements used in this study were taken in the Sea of Japan, at a location 3 km off the Yura fishery harbor, where the water depth is about 43 meters (Fig.\ref{Yura}). The data set consists of $1.08 \times10^5$ samples at a sampling frequency of $1$ Hz \cite{Liu2000}. The measuring device is an ultrasonic wave gauge manufactured by Kaijo Sonic Corporation. Mori et al. \cite{Mori2002a,Mori2002b,Mori2002c,Liu2000} have published several studies with data from the location.

\begin{figure}[H]
  \begin{center}
    \includegraphics[width=.49\textwidth]{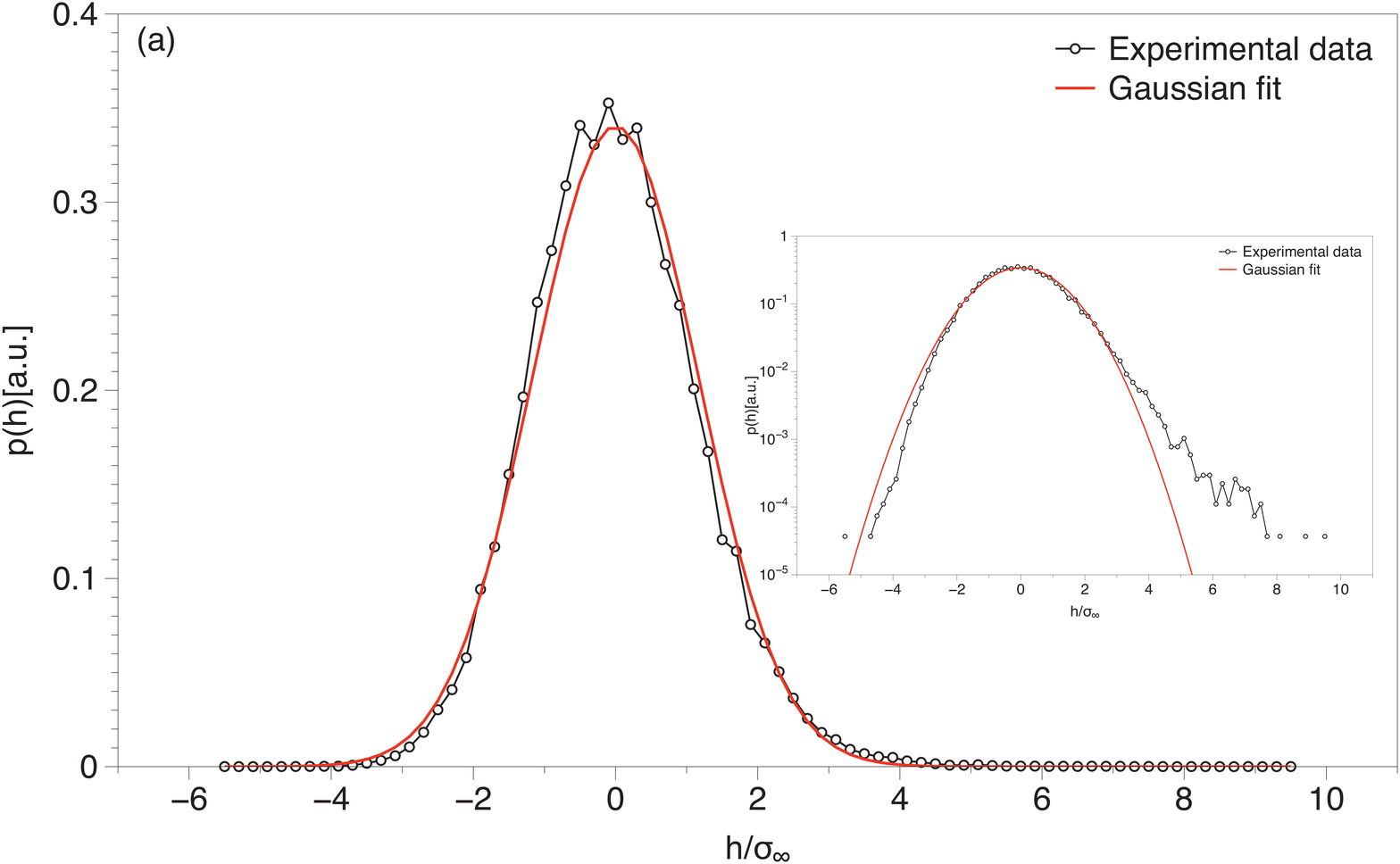}
     \includegraphics[width=.49\textwidth]{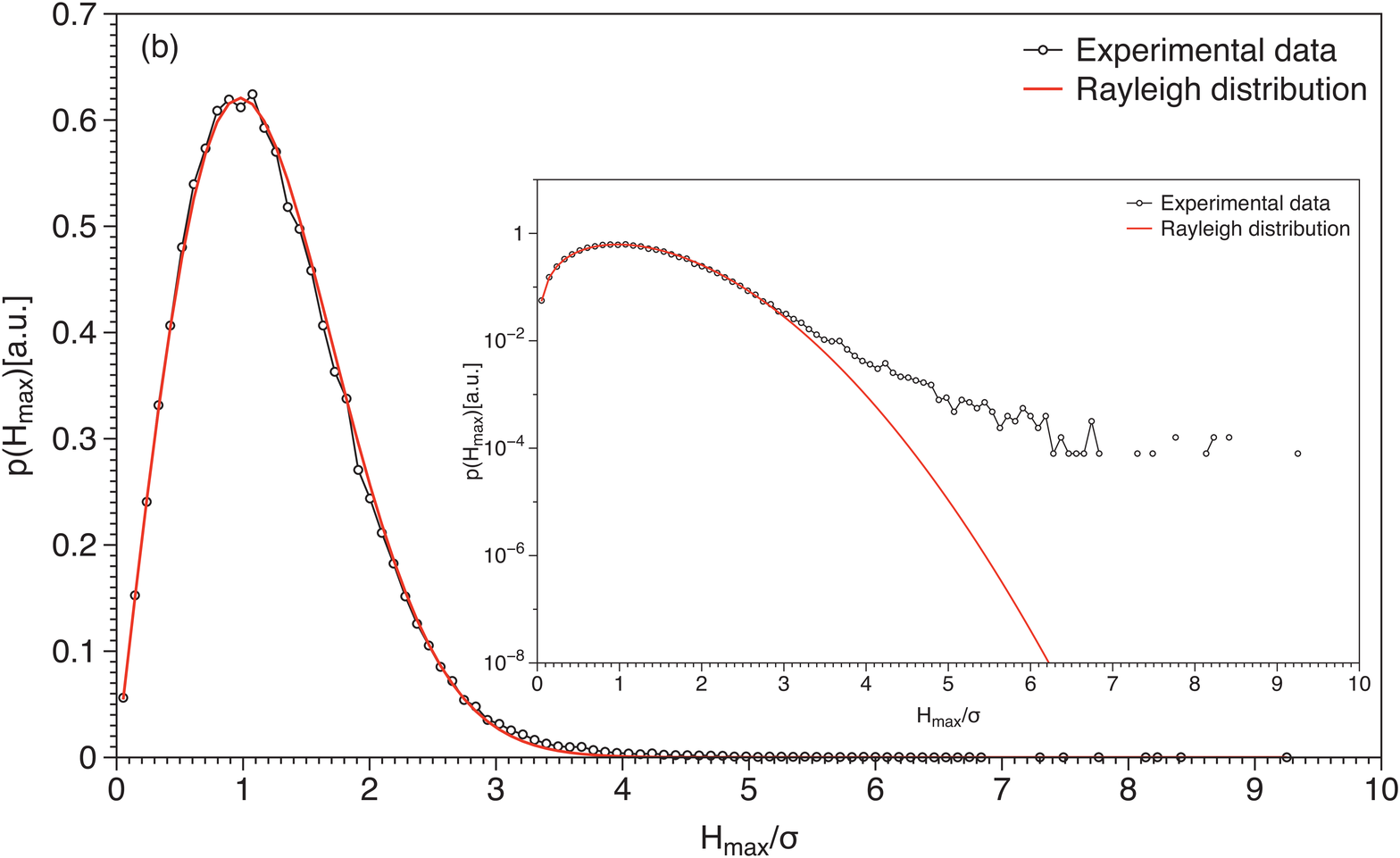}
     
  \end{center}
 \caption{(a) Probability density function of surface elevation data and a Gaussian fit (solid line). (b) Probability density function of surface elevation maxima in linear and logarithmic scale, and a Rayleigh fit (solid line).}
 \label{pdfs}
\end{figure}

First we will compare statistical properties of the experimental surface elevation data with Gaussian and Rayleigh distributions. Fig.\ref{pdfs} a shows the surface elevation distribution of the time-series and a best fitting Gaussian distribution. Fig.\ref{pdfs} b gives the peak value distribution, i.e. the distribution of the local surface elevation maxima, and a best fitting Rayleigh distribution. Wave height is given in units of $\sigma_\infty$ which is defined according to: $\sigma^2_\infty=\lim_{\tau \to \infty}<h(\tau)^2>.$ It can be observed that the Gaussian and the Rayleigh distributions characterize the overall statistics quite well. However, in a semi-logarithmic presentation for the large events remarkable deviations are found. The Gaussian and Rayleigh distributions underpredict the probabilities of such large events by very large factors.

\begin{figure}[H]
  \begin{center}
    \includegraphics[width=17 cm, height=6 cm]{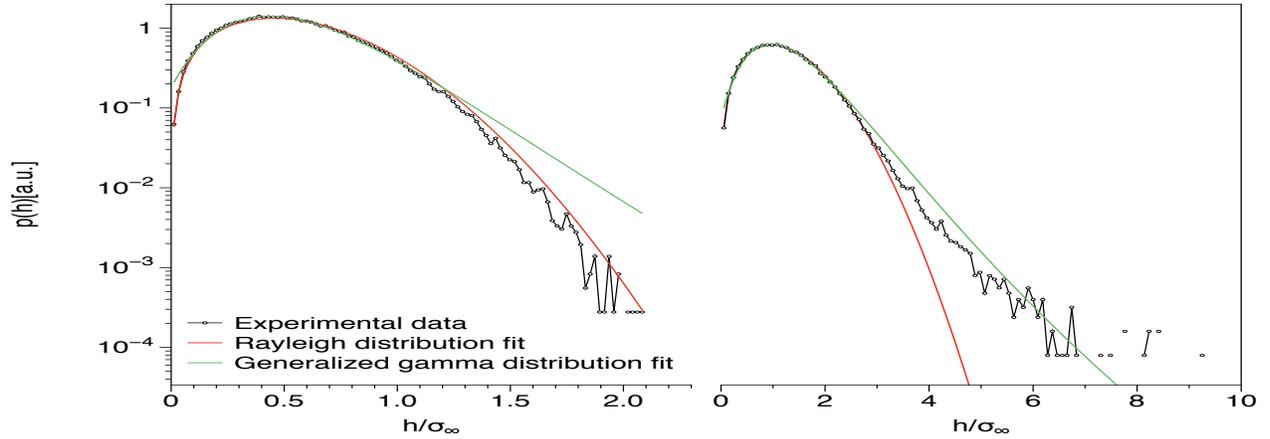}
    
  \end{center}
 \caption{Probability density functions of local surface elevation maxima. Fino data (a) and Yura (b). }
 \label{pdf}
\end{figure}

To gain further insight if indeed the observed 'fat tails' can be related to the occurrence of rogue waves, another data set was analyzed, for which it is expected that no rogue waves are included: We picked a wave data recording from the North sea based Fino research platform, which had also been sampled with a frequency of 1 Hz. The corresponding peak value distributions for the Yura and the Fino data are shown in Fig.\ref{pdf}. The Fino data follow nicely the Rayleigh distribution, whereas the Yura data clearly deviate from it. The Yura data can be fitted quite well by a generalized gamma distributions 
\begin{equation}
p(h)=\frac{\nu}{\Gamma(\alpha/\nu)} \frac{|\beta|^{2\alpha}}{|h+T_0|^{\alpha+1}} exp\{-(\frac{\beta^2}{h+T_0})^{\nu}\},
\label{gamma}
\end{equation}
with the parameters $\alpha=0.34$, $\nu=3.34$, $\beta=6.54$ and $T_0=3.26$ \cite{GR Jafari2006,Shayeganfar2012,Anvari2013}. Although the data follow gamma distribution quite well, on may also see a transition at $h=4$ from a Rayleigh distribution to an exponential tail ($e^{-h}$) for large events. \\

As we will focus in this paper on large events, or, respectively rogue waves, we will restrict in the following our study to the Yura data.  All data will be normalized to zero mean and a standard deviation of one ($\sigma = 1 m$).


\section{Markov properties, multi-scale analysis, and the technique to derive Fokker-Planck equations}

Complex systems typically have hierarchical and non-trivial structures on different scales. Such systems are often more successfully described as multi-scale processes, rather than processes in time or space \cite{Friedrich1997,Waechter2004,Jafari2003,Tabar2006}. In the following we will employ these ideas to the data described above. The starting point is the surface elevation $h(t)$. To resolve the scale dependent complexity,  increments, defined as $h_{\tau}(t):= h(t+\tau)-h(t)$, are taken as measure of the scale dependent structure. Here $\tau$ is the increment time interval which is used to analyze different time scales. 

Complete information about the complexity of the height structure $h(t)$ would be provided by the general joint pdfs $p(h_n, \tau_n;. . .;h_1, \tau_1;h_0, \tau_0)$ of increments at different scales $\tau$ \cite{Friedrich2011}. Here and in the following we use the simplified notation of $h_{\tau_n}(t) := h_n(t)$. In terms of conditional pdfs, the joint pdfs can be written as 
\begin{eqnarray}
\nonumber
&p(h_n, \tau_n;. . .;h_1, \tau_1;h_0, \tau_0)=\\
\\
 \nonumber
 & p(h_n,\tau_n | h_{n-1},\tau_{n-1}; ... ;h_0,\tau_0).p(h_{n-1},\tau_{n-1} | h_{n-2},\tau_{n-2}; ... ;h_0,\tau_0) ... p(h_1,\tau_1|h_0,\tau_0).p(h_0,\tau_0)  .
\label{general_pdf}
\end{eqnarray}
Because of the involved scales, the dimension of the joint pdf is very high. Thus in general it is very hard to estimate it from time series data directly. However, the description and computation can be highly simplified if the conditional pdfs obey the condition
\begin{equation}
p(h_ n, \tau_n|h_{n-1}, \tau_{n-1}; ... ; h_1,\tau_1 ; h_0,\tau_0)=p(h_ n, \tau_n|h_{n-1}, \tau_{n-1}).
\label{3}
\end{equation}
This property (\ref{3})is nothing else than the Markov property for a process of $h_{\tau}$ evolving in $\tau$, and reduces the multi conditioned pdf to a simple conditioned one. Only the nearby data, $(h_{n-1}, \tau_{n-1})$, is relevant to the probability of finding the system at a particular state $h_n$ in scale $\tau_n$. Whereas the multi conditioned pdf involves the information of the height structure at many instances, the simple conditioned pdf $p(h_ n, \tau_n|h_{n-1}, \tau_{n-1})$ depends only on three instances, namely $h(t), h(t+\tau_{n-1})$ and $h(t+\tau_{n})$. In this sense the Markov property is a three point closure of the general n-scale pdf.

\begin{equation}
p(h_n,\tau_n ; ... ; h_{1},\tau_{1};h_0,\tau_0)=p(h_n,\tau_n | h_{n-1},\tau_{n-1}) ... p(h_1,\tau_1|h_0,\tau_0)p(h_0,\tau_0).
\end{equation}

It is well known, that a given process fulfils the stochastic Markov property only for an infinite Markov-Einstein (ME) length scale \cite{Einstein1905}, which is the minimum length scale over which the data can be considered as a Markov process \cite{Risken1989}. For Markov processes the conditional probability density fulfils a master equation which can be put into the form of a Kramers-Moyal expansion, \cite{Kolmogorov1931,Risken1989}
\begin{equation}
-\tau \frac{\partial}{\partial \tau}\;p(h, \tau|h_0, \tau_0)\,=\sum_{k=1}^{\infty}(-\frac{\partial}{\partial h})^k D^{(k)}(h,\tau)p(h,\tau|h_0,\tau_0),
\end{equation}
where the Kramers-Moyal coefficients, $D^{(k)}(h,\tau)$, are defined as limits $\Delta \tau \rightarrow 0$ of the conditional moments
\begin{equation}
D^{(k)}(h,\tau)=\lim_{\Delta \tau \to 0}M^{(k)}(h,\tau,\Delta \tau),
\label{6}
\end{equation}
\begin{equation}
M^{(k)}(h,\tau,\Delta \tau)=\frac{\tau}{k!\Delta \tau}\int_{-\infty}^{+\infty}(\tilde{h}-h)^kp(\tilde{h},\tau - \Delta \tau | h,\tau)d\tilde{h}.
\label{7}
\end{equation}
For a general stochastic process, all Kramers-Moyal coefficients are non-vanishing. If, however, the fourth coefficient $D^{(4)}(h,\tau)$ vanishes, Pawula's theorem \cite{Risken1989} states that only the coefficients $D^{(1)}(h,\tau)$ and $D^{(2)}(h,\tau)$ are non-zero. $D^{(1)}(h,\tau)$ is called the drift-term and $D^{(2)}(h,\tau)$ the diffusion-term. For this case the Kramers-Moyal-expansion reduces to a Fokker-Planck-equation of the form
\begin{equation}
-\tau \frac{\partial}{\partial \tau}\;p(h_ \tau, \tau|h_0, \tau_0)\,=\{-\frac{\partial}{\partial h_ \tau} D^{(1)}(h_ \tau, \tau)+ \frac{\partial^2}{\partial h_ \tau^2} D^{(2)}(h_ \tau, \tau)   \}p(h_ \tau, \tau|h_0,\tau_0).
\end{equation}
Also the probability density $p(h_{\tau},\tau)$ has to obey the same equation:
\begin{equation}
-\tau \frac{\partial}{\partial \tau}\;p(h_ \tau, \tau)\,=\{-\frac{\partial}{\partial h_ \tau} D^{(1)}(h_ \tau, \tau)+ \frac{\partial^2}{\partial h_ \tau^2} D^{(2)}(h_ \tau, \tau)   \}p(h_ \tau, \tau).
\label{fokker-planck}
\end{equation}
Next we analyze the above mentioned data, and show how to ensure Markov properties, how to derive Fokker-Planck equations, and how to interpret the results.

\section{Analysis and Results}
\subsection{Markov properties}
\begin{figure}[H]
  \begin{center}
    \includegraphics[width=.5\textwidth]{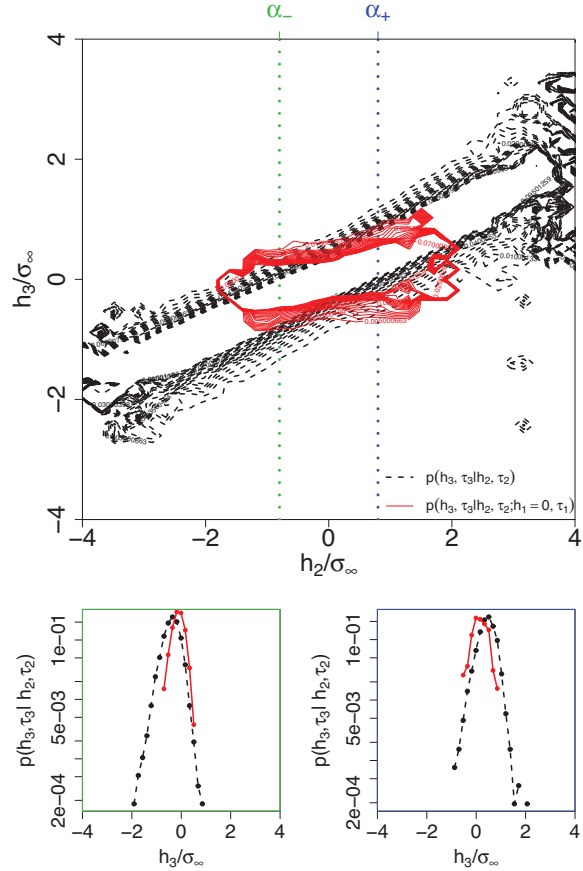}
  \end{center}
 \caption{Contour plots of the conditional pdfs $p(h_3,\tau_3 | h_2,\tau_2)$ (dashed black lines) and $p(h_3,\tau_3 | h_2,\tau_2 ; h_1=0,\tau_1)$ (solid red lines) for $\tau_1=19, \tau_2=2\tau_1, \tau_3=3\tau_1$. Cuts through the conditional pdfs for $h_2=-\infty$ and $h_2=+\infty$. Solid red line: $p(h_3,\tau_3|
 h_2,\tau_2;h_1=0,\tau_1)$, black dashed line: $p(h_3,\tau_3|h_2,\tau_2)$.}
 \label{cpdf}
\end{figure}

We first have to show for the conditional pdfs that the Markov property is fulfilled. This is in general difficult for higher conditional pdfs, but it is possible for $n=3$. The relation
\begin{equation}
p(h_3,\tau_3 | h_2,\tau_2;h_1,\tau_1)=p(h_3,\tau_3 | h_2,\tau_2) 
 \label{Mar}
\end{equation}
should hold for any value of $\tau_2$ in the interval $\tau_1 <\tau_2 < \tau_3$. In fig. \ref{cpdf} we compare double and single conditional pdfs for the present experimental data to check this requirement of equation \ref{Mar}.


\begin{figure}[H]
  \begin{center}
    \includegraphics[width=.5\textwidth]{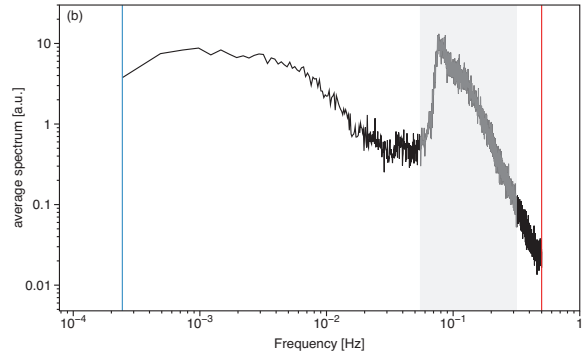}
  \end{center}
 \caption{Fourier spectrum for the water surface elevation data. The highlighted range includes a sharp peak and indicates a dominant frequency leading to long range correlation in the data. }
 \label{spect}
\end{figure}

\begin{figure}[H]
  \begin{center}
    \includegraphics[width=.49\textwidth, height=6 cm]{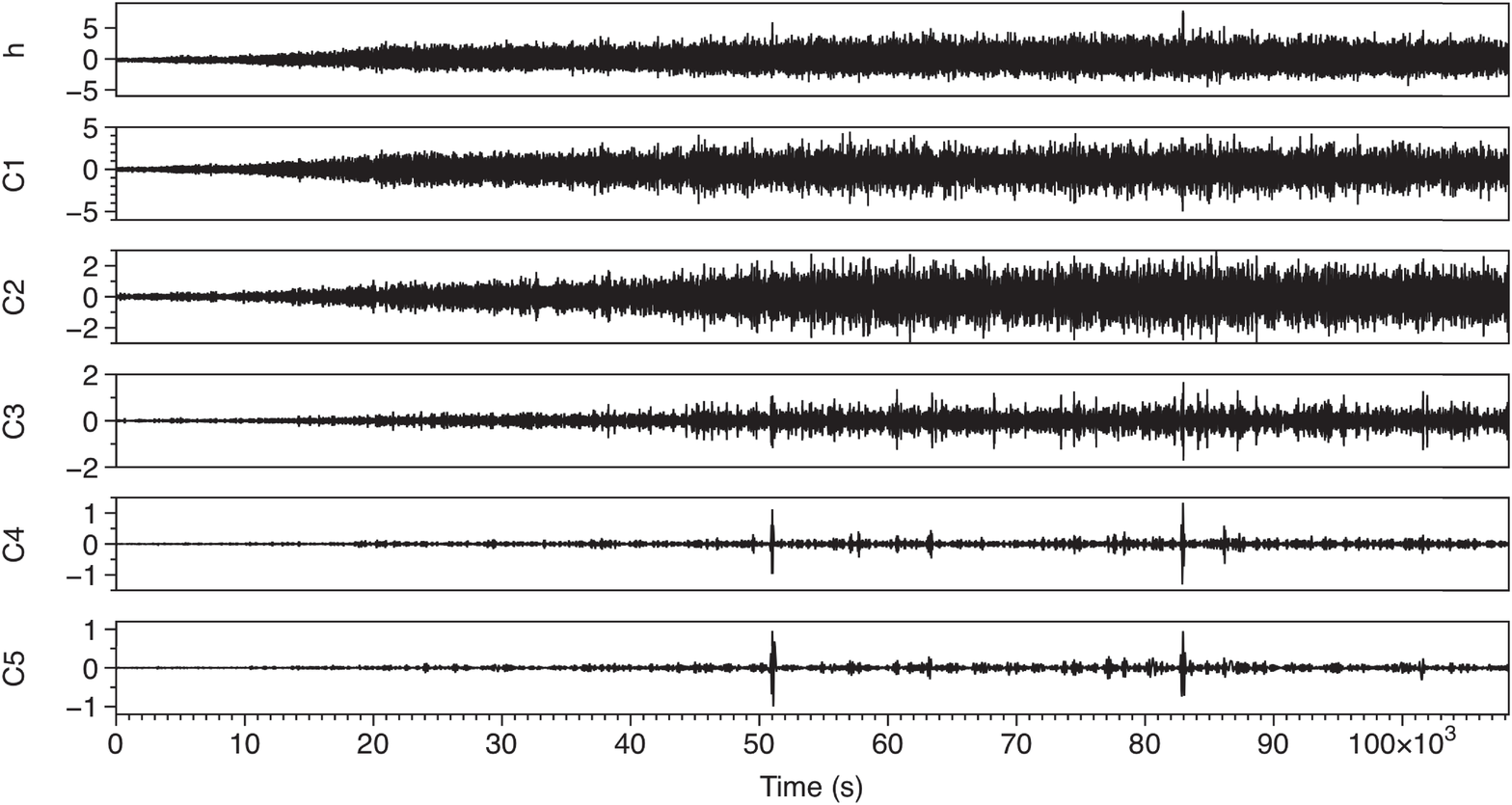}
    \includegraphics[width=.49\textwidth, height=6 cm]{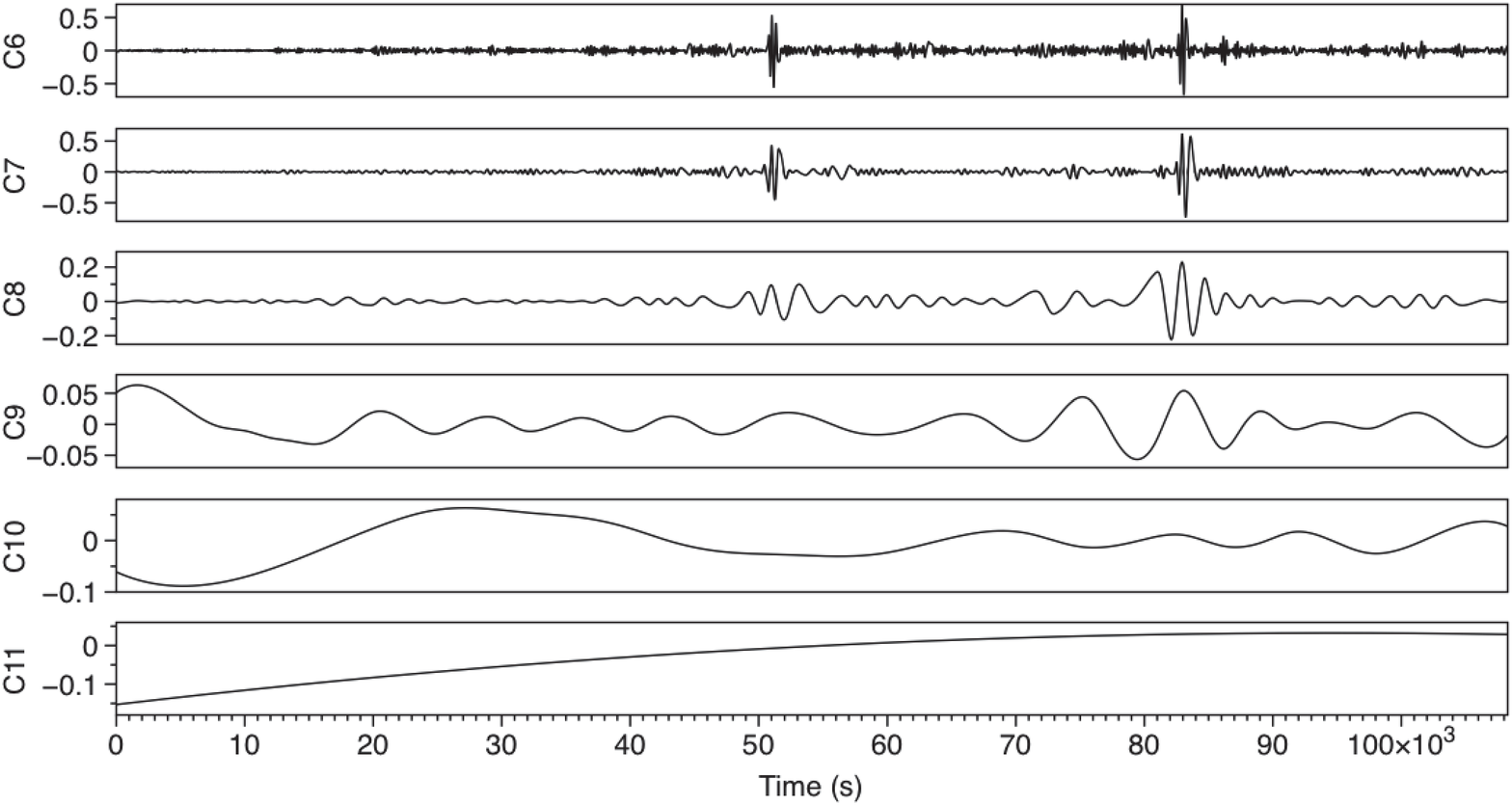}
  \end{center}
 \caption{The IMFs (C1 to C11) for the data $h(t)$ derived from Empirical Mode Decomposition.}
 \label{emd}
\end{figure}

\begin{figure}[H]
  \begin{center}
    \includegraphics[width=.75\textwidth]{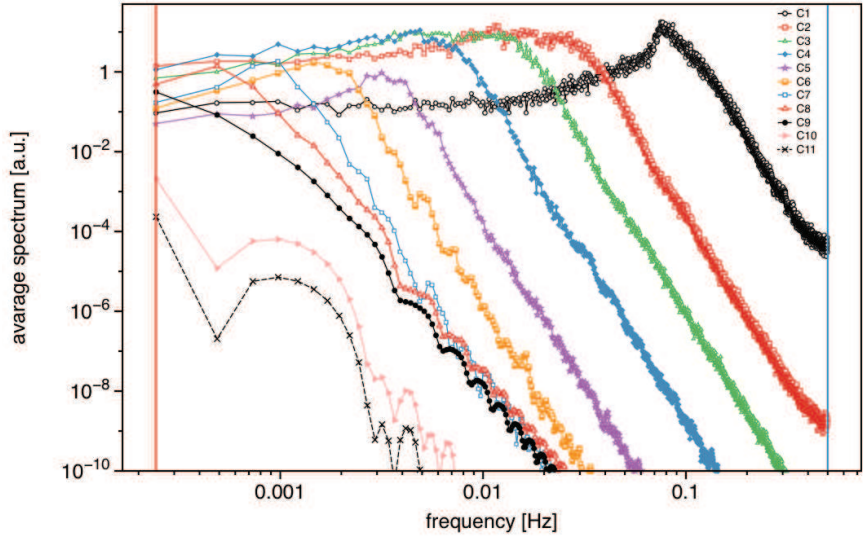}
  \end{center}
 \caption{Fourier spectra for all IMFs (C1 to C11). The higher frequency components are dominantly confined to C1.}
 \label{EMD_spect}
\end{figure}

Obviously the double and single conditional pdfs deviate substantially from each other, so the necessary condition for a Markov process is not fulfilled, which makes the direct application of the approach described above impossible. To understand the situation better, and to possibly identify the origin of the non-Markovian characteristics, we first had a look at the spectral characteristics of our data set. Fig. \ref{spect} shows the Fourier spectrum of our data, and it is clearly visible that there is a sharp peak indicating a strong high-frequency component in the data. Obviously this peak is related to the background wave state, and indicates that a comparatively small frequency band plays an important role in the process. Such a narrow frequency band plausibly suggests that in the signal there is a long range correlation in time and space, which in turn may be in conflict with the Markov properties needed for the present analysis.  Here one should note that correlation and non-Markovian memory are not the same. Correlations caused by the drift term $D^{(1)}$ will have no memory, whereas correlated noise in the corresponding Langevin equation (within the diffusion term) causes memory effects \cite{Risken1989}.
To cope with these non-Markovian effects, we first tried a decomposition of our wave data in Fourier modes. After filtering out some of the modes responsible for the high frequency peak we already saw a better fulfilment of the Markov conditioned (\ref{3}). Best results were obtained by the application of Empirical Mode Decomposition (EMD). Details of EMD can be found in \cite{huang1998a,huang1998b,huang1999a,Battista2007,Nunes2003,huang2003,huang1999b}. EMD aims at breaking down the original signal into a hierarchy of Intrinsic Mode Functions (IMF) that separate the different frequency components of the signal by counting maxima and zero crossings. Fig.\ref{emd} shows the results of the EMD.
By looking at the Fourier spectra of the IMFs, which are shown in Fig.\ref{EMD_spect}, it turns out that the effect of the dominant frequency is largely confined to the first IMF, C1, only. We thus hypothesized that this decomposition method could provide a promising new filtering opportunity to eliminate the non-Markovian contribution, and to consequently make the stochastic approach possible.

\begin{figure}[H]
  \begin{center}
    \includegraphics[width=.49\textwidth, height=6 cm]{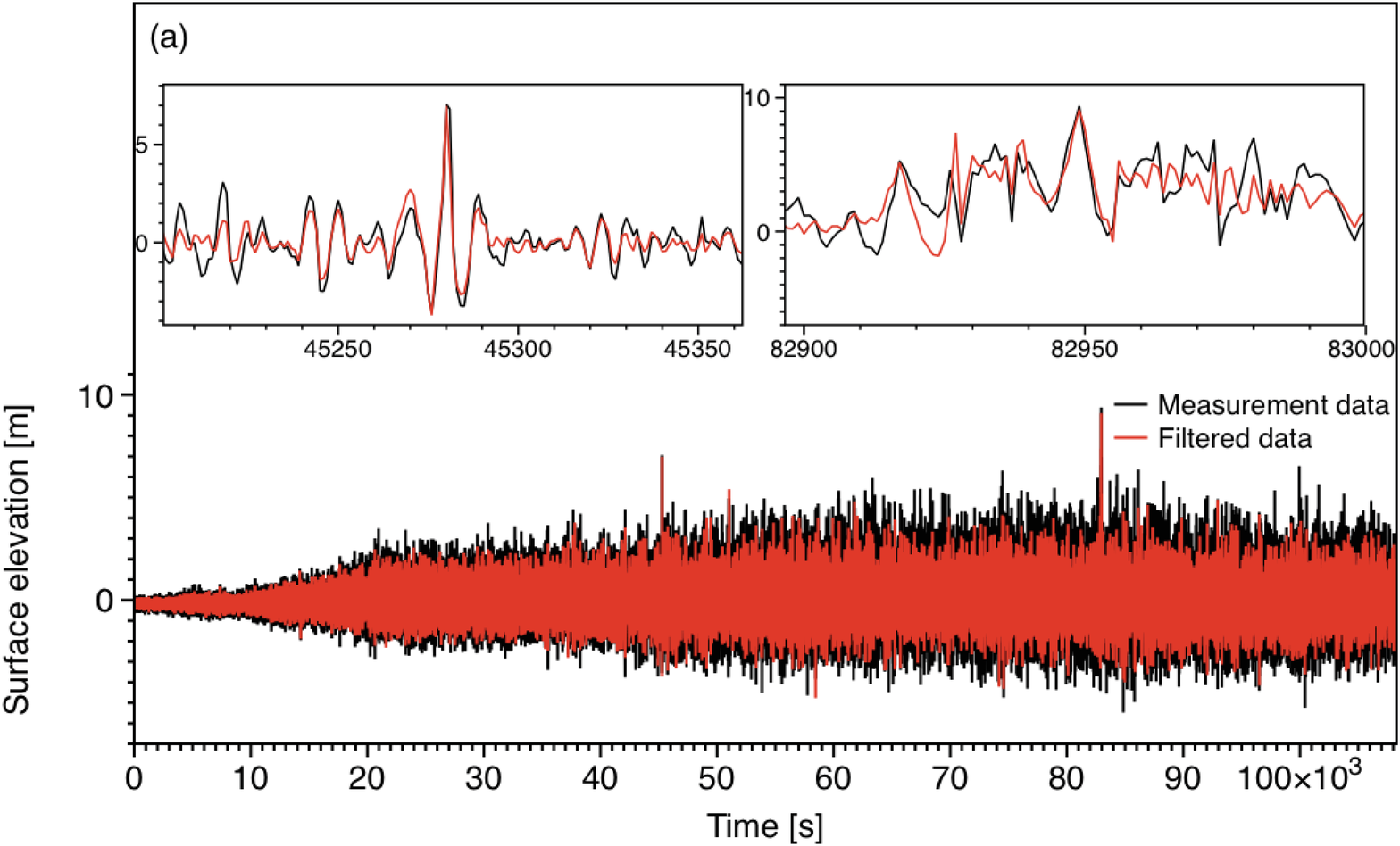}
    \includegraphics[width=.49\textwidth, height=6 cm]{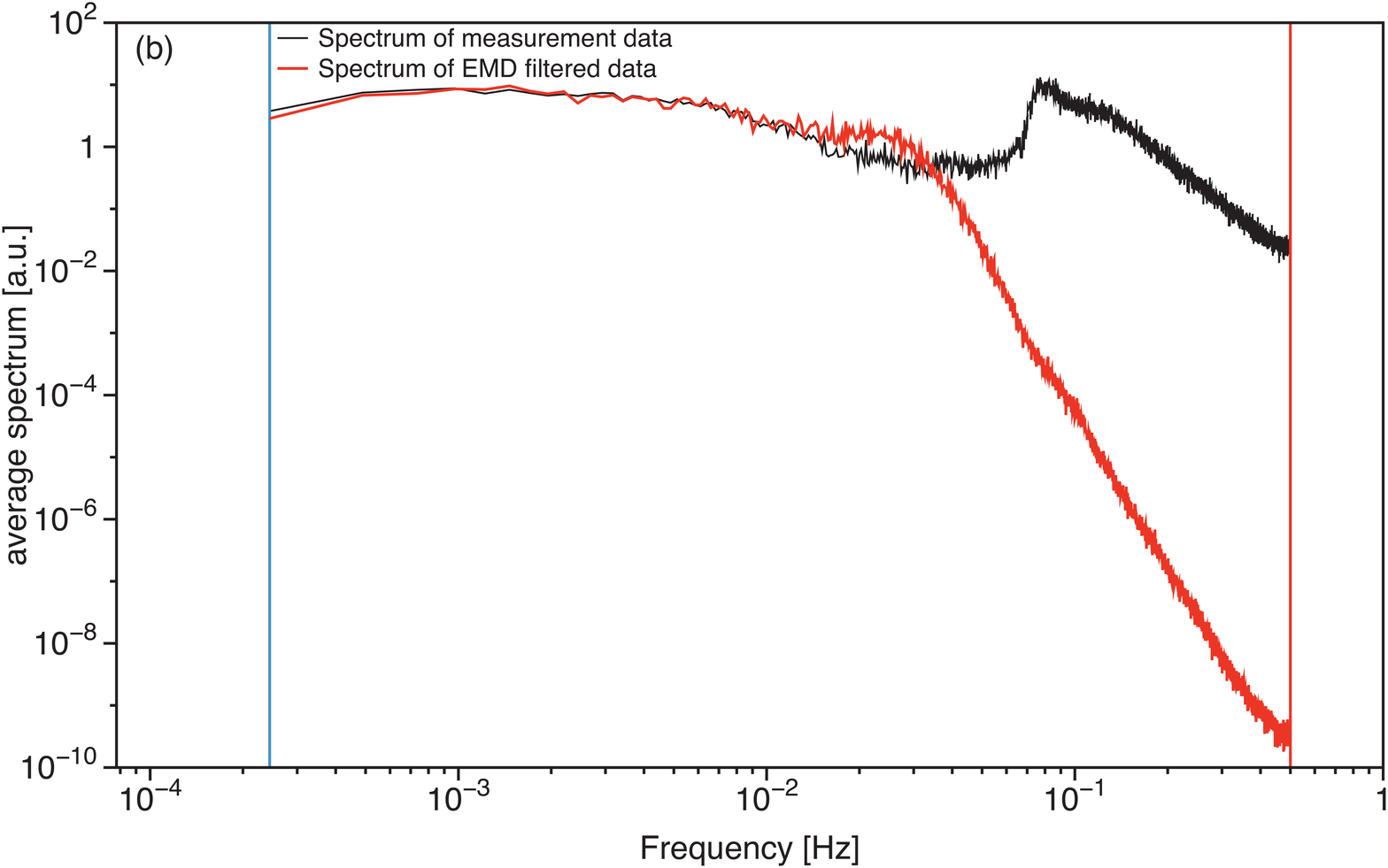}
  \end{center}
 \caption{(a) Original time-series data (black) compared to EMD filtered data (red), (b) Spectrum of original data (black) and filtered data (red).}
 \label{hht_filter1}
\end{figure}

We therefore removed component C1 from the signal and reconstructed the data by adding C2 up to C11. The resulting time series and frequency spectra are shown in Fig.\ref{hht_filter1}a and Fig.\ref{hht_filter1}b, respectively. First, it turns out that the filtering approach has successfully removed the high-frequency component from the signal. Remarkably, it seems we still have all the extreme wave events, or rogue waves, in the reconstructed data, as can be seen from Fig.\ref{hht_filter1}a . At first sight this seems surprising, since the filtering appears to have removed the dominant background wave component. However, the result might in fact also be a confirmation of the hypothesis that rogue waves are indeed caused by nonlinear breather type dynamics of the wave field: Breather states have the remarkable property of phase jumps. For example, the Peregrine breather is characterized by a phase jump of $2 \pi$ across its spatial coordinate. In terms of physical characteristics of the wave field in the neighbourhood of a breather state, one would thus indeed expect significant local changes of wavelength and wave frequency. This is indeed confirmed for the large waves given in the present data set: from the EMD results it turns out that the wave dynamics in the phases of rogue events is mainly captured by higher IMFs, here say C3 to C8, i.e. IMFs corresponding to slightly shifted underlying numbers of waves.

\begin{figure}[H]
  \begin{center}
    \includegraphics[width=.49\textwidth, height=12 cm]{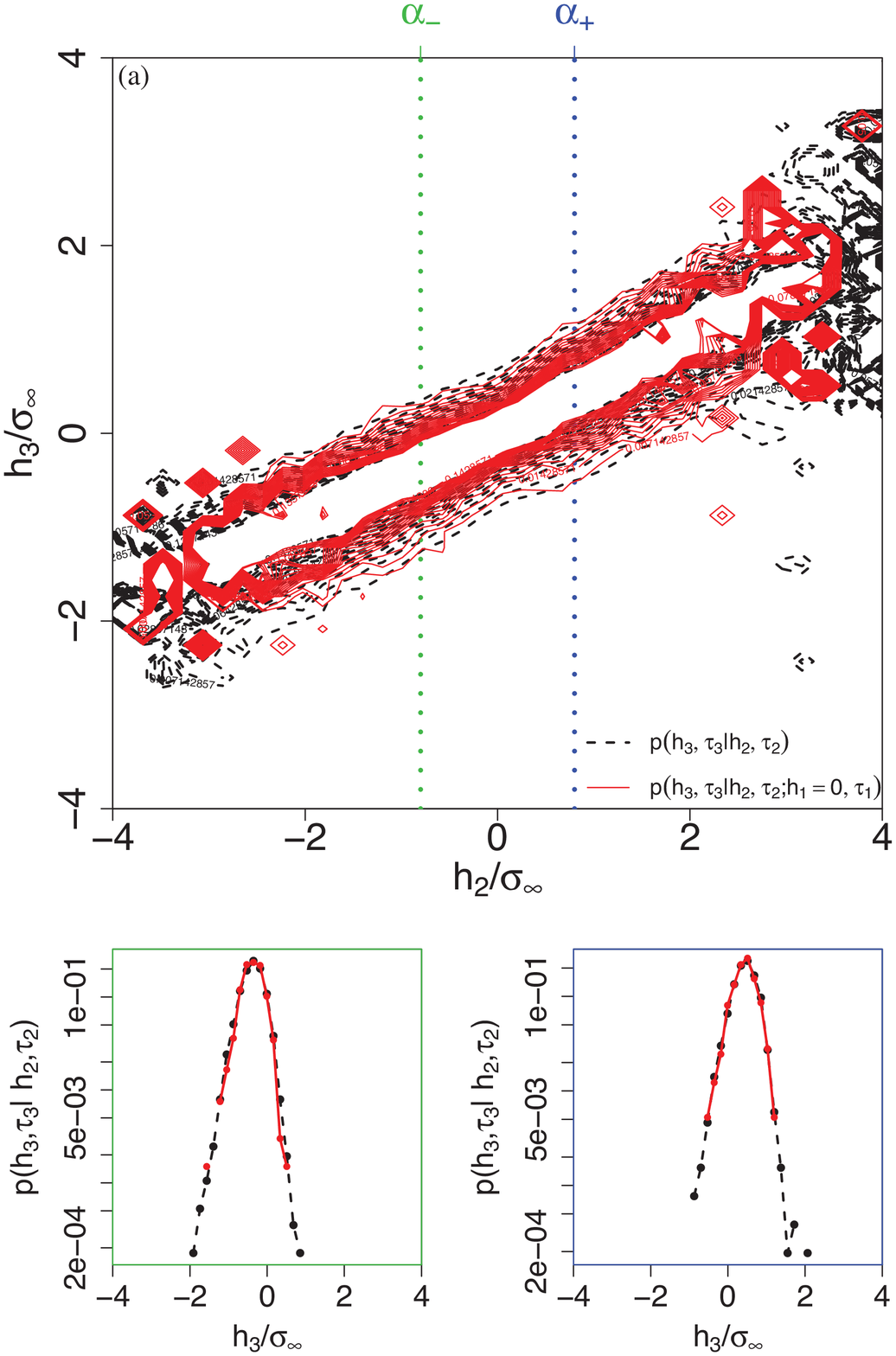}
    \includegraphics[width=.49\textwidth, height=12 cm]{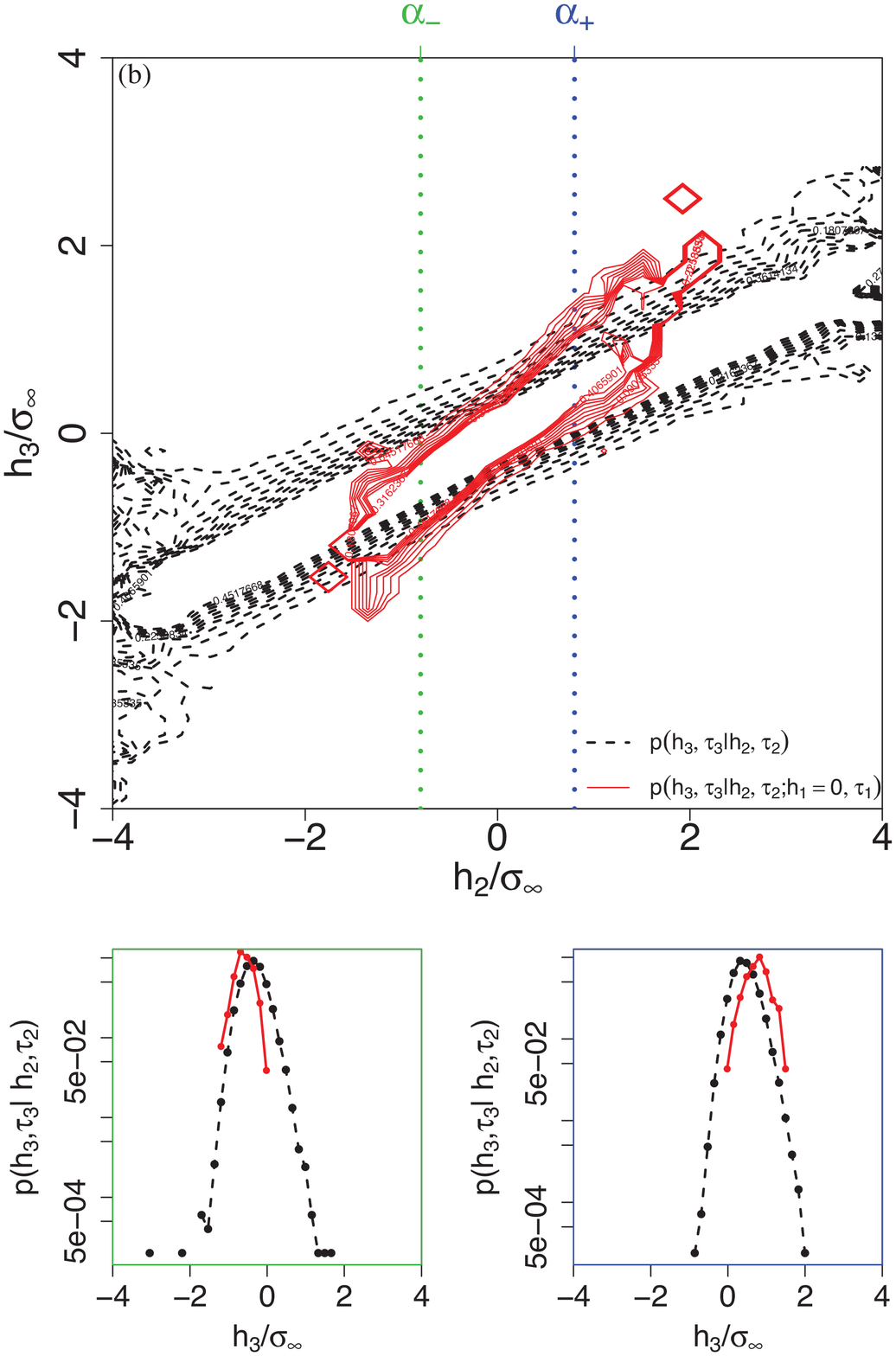}
  \end{center}
 \caption{As figure 3, (a) contour plots of the conditional pdfs for data with C1 filtered out, and (b) for C1 alone. $p(h_3,\tau_3 | 
h_2,\tau_2)$ (black lines) and $p(h_3,\tau_3 | h_2,\tau_2 ; h_1=0,\tau_1)$ (red lines).}
 \label{cpdf_hht}
\end{figure}

For our original approach the EMD was successful to remove the dominant frequency. To see if indeed the memory has been removed, we now test for Markovian properties of the new data set. In Fig.\ref{cpdf_hht}a contour plots of $p(h_3,\tau_3 | h_2,\tau_2;h_1,\tau_1)$ and $p(h_3,\tau_3 | h_2,\tau_2)$ have been superposed. The proximity of corresponding contour lines yields highly satisfying evidence for Markovian properties for the chosen set of scales. Additionally, two cuts through the conditional probability densities are provided for fixed values of $h_2$, further supporting this result. For completeness, Fig.\ref{cpdf_hht}b shows the same analysis for the first IMF. It becomes highly plausible, comparing this to the previous analysis based on the full data, that the first IMF is mainly responsible for the non-Markovian properties of the original overall process.

\begin{figure}[H]
  \begin{center}
    \includegraphics[width=.8\textwidth]{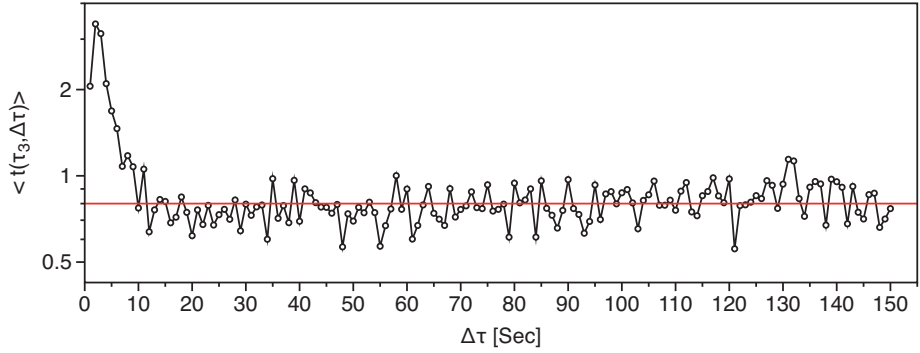}
  \end{center}
 \caption{The expectation value $<t(\tau_3,\Delta \tau)> $ plotted as a function of $\Delta \tau$.}
 \label{wil}
\end{figure}

To quantify the Markov properties we also performed a Wilcoxon test for the scales $\tau_1$, $\tau_2$ and $\tau_3$, for details of the method see \cite{Renner2001b}. The normalized expectation value $<t>$ of the number of inversions of the conditional velocity increments $h_3|_{h_2}$ and $h_3|_{h_2,h_1}$ was calculated. If Markovian properties exist, $<t>$ has a value of $\sqrt{2/\pi}\approx 0.8$. The expectation value $<t>$ is a function of the surface elevation increment $h_1$ and the length scales $\tau_1$, $\tau_2$ and $\tau_3$. In order to reduce the number of parameters, we chose $h_1$ to be zero and the difference $\tau_3-\tau_2$ and $\tau_2-\tau_1$ to be equal: $\Delta \tau=\tau_3-\tau_2=\tau_2-\tau_1$. In figure \ref{wil}, $<t>$ is plotted as a function of $\Delta \tau$ for scale $\tau_3$. The Wilcoxon test indicates that Markov properties exist, as long as $\Delta \tau$ is chosen to be larger than at least $14$ seconds. Note, $\Delta \tau = 14s$ is here the Markov-Einstein length. 
 
\subsection{Kramers-Moyal coefficients and Fokker-Planck equations}

\begin{figure}[H]
  \begin{center}
    \includegraphics[width=.49\textwidth,height=6 cm]{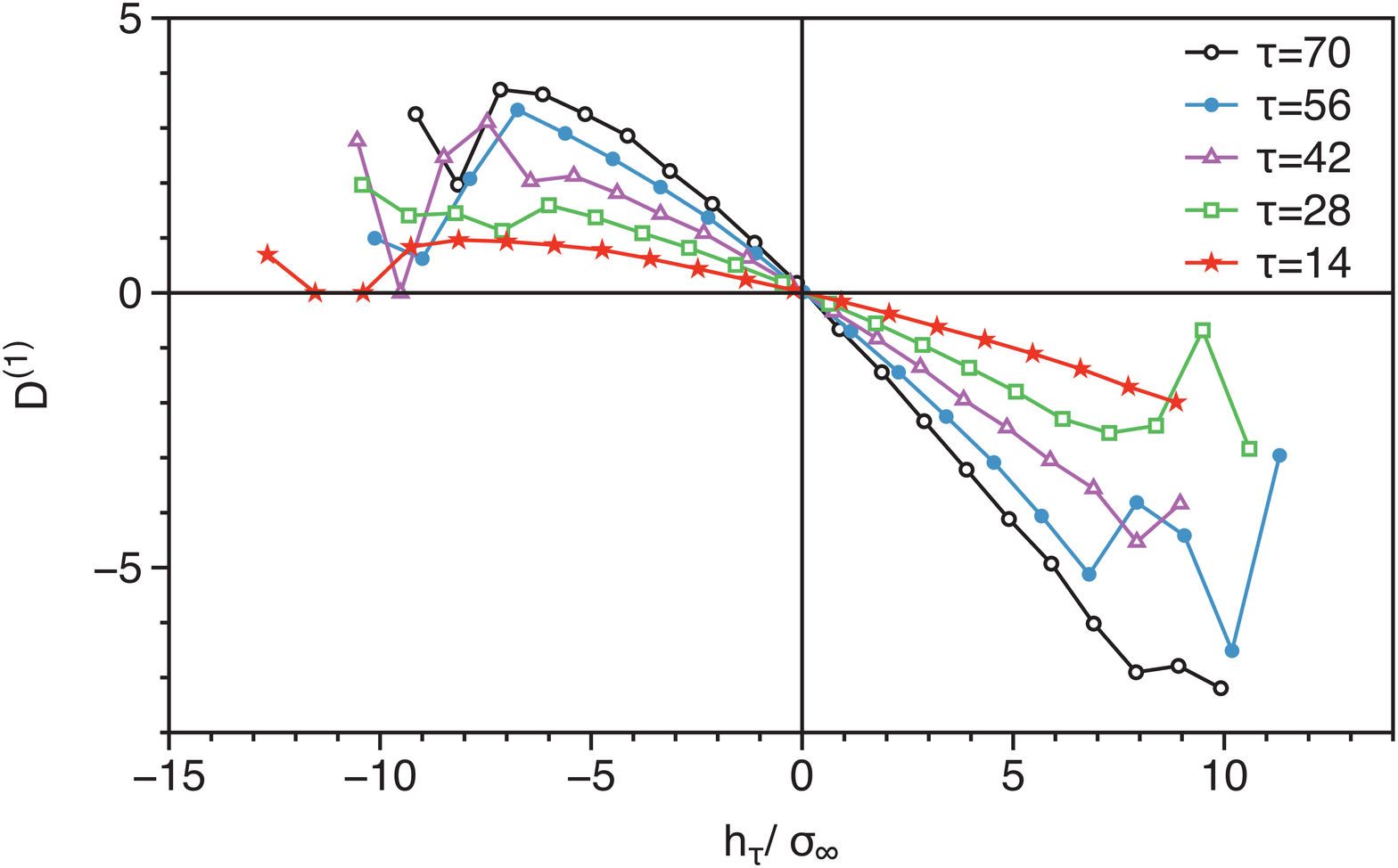}
    \includegraphics[width=.49\textwidth,height=6 cm]{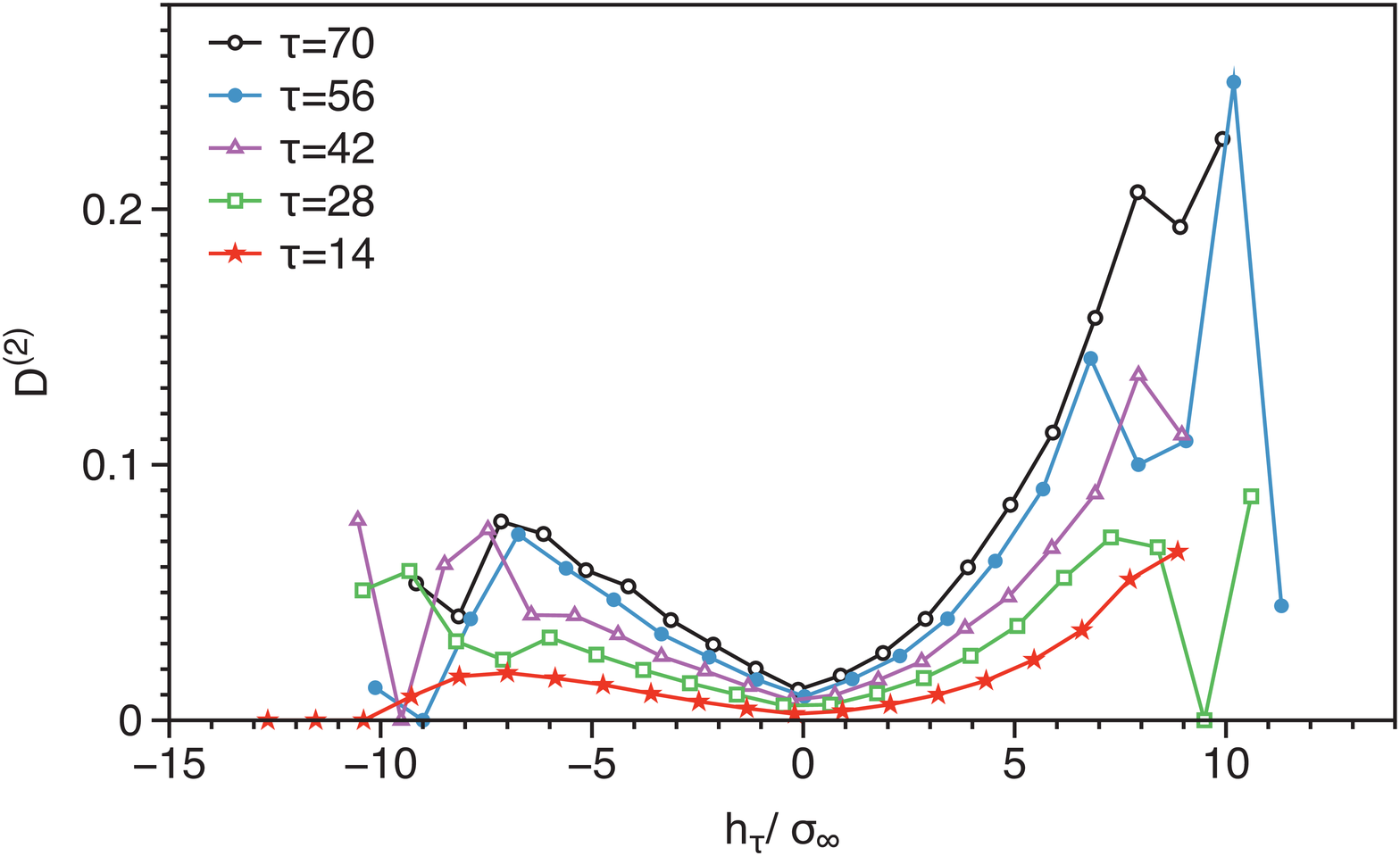}
  \end{center}
  \caption{Estimated drift ($D^{(1)}(h_\tau,\tau)$) and diffusion ($D^{(2)}(h_\tau,\tau)$) coefficients for different scales ($\tau$).}
  \label{D1_D2}
\end{figure}

As we have shown that the EMD filtered wave date show Markov properties in scale $\tau$ we can proceed to calculate, according to equation (\ref{7}), the conditional moments, $M^{(k)}(h,\tau,\Delta\tau)$ using the empirically determined conditional probability density functions. The Kramers-Moyal coefficients can be obtained in the limit $\Delta\tau \to 0$ using equation (\ref{6}). Fig. (\ref{D1_D2}) shows the resulting drift and diffusion coefficients $D^{(1)}$ and $D^{(2)}$ functions, respectively. The data analysis has been conducted for a number of different time-increments, and the results indicate the multi-scale nature of the process. 

\begin{figure}[H]
  \begin{center}
    \includegraphics[width=.49\textwidth]{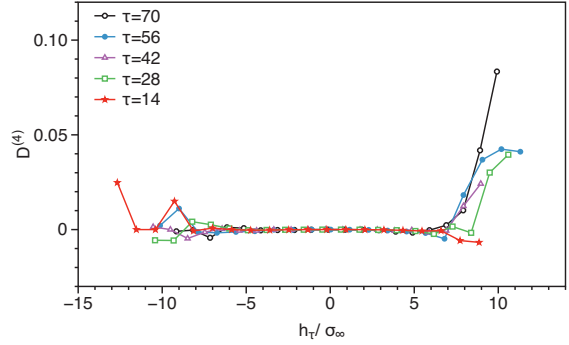}
  \end{center}
 \caption{Estimated $D^{(4)}$ coefficient for different scales.}
 \label{D4}
\end{figure}

As mentioned, the magnitude of the fourth Kramers-Moyal coefficient, $D^{(4)}$, is important. If $D^{(4)}$ can be taken as zero, the whole scale dependent complexity can be described by a comparatively simple Fokker-Planck equation, what means that Gaussian distributed and delta correlated noise is present in the stochastic scale process. Fig.\ref{D4} shows the results of the corresponding analysis, and indeed the reduction to a Fokker-Planck equation seems highly plausible.

\begin{figure}[H]
  \begin{center}
    \includegraphics[width=.69\textwidth]{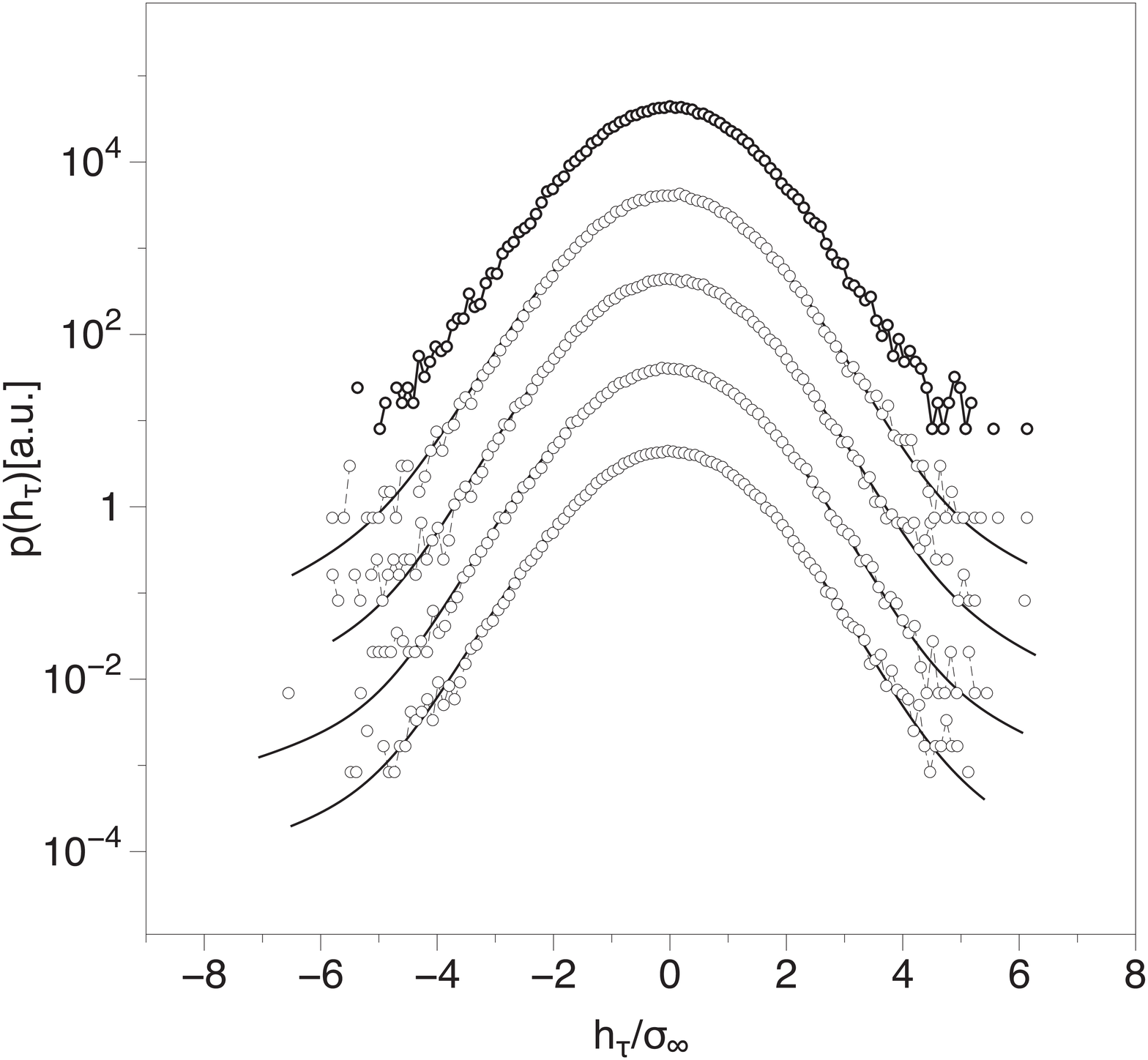}
  \end{center}
 \caption{Comparison of the numerical solution of Fokker-Planck equation (solid lines) for the pdfs $p(h_\tau,\tau)$ with the pdfs obtained directly from the data (symbols). The scales $\tau$ are (from top to bottom): $\tau=70, 56, 42, 28$ and $14$s. The pdf at largest scale was parametrized and used as initial condition for the iteration of the Fokker-Planck equation. Curves are shifted in the vertical direction for clarity of presentation. }
 \label{fp}
\end{figure}

With these results Fokker-Planck equations for the relevant pdfs (Eq.\ref{fokker-planck}) can be formulated. $D^{(1)}$ and $D^{(2)}$ completely determine the equations, and by themselves give a characterization of the multi-scaled process. Although the analysis leading to this conclusion was already quite plausible, to gain further confidence, we compared solutions of the resulting Fokker-Planck equations and the resulting pdfs with the corresponding distributions obtained directly from the data. The solution of the Fokker-Planck equation for small $\Delta \tau$ is given by \cite{Risken1989}
\begin{equation}
	p(h,\tau+\Delta\tau|h^{\prime},\tau)=\frac{1}{\sqrt{4\pi D^{(2)}(h)\Delta\tau}}\times\exp{[\frac{-(h-h^{\prime}-D^{(1)}(h)\Delta\tau)^2}{4D^{(2)}(h)\Delta \tau}]}.
	\label{propagator}
\end{equation}
In order to obtain pdfs for larger steps, we use the Chapman-Kolmogorov equation 
\begin{equation}
	p(h,\tau+\Delta\tau)=\int p(h,\tau+\Delta\tau|h^{\prime},\tau)p(h^{\prime},\tau)dh^{\prime}.
	\label{chapman-kolmogrov}
\end{equation}
Equation (\ref{chapman-kolmogrov}) is a direct consequence of the Markov condition, too. By iterating this procedure, we finally obtain pdfs on different scales. Figure \ref{fp} compares the solution of Fokker-Planck equations for the pdf $p(h_\tau,\tau)$ with the empirically estimated pdfs from the data set. The comparison proves strikingly that the Fokker-Planck equation accurately describes the evolution of $p(h_\tau,\tau)$ in $\tau$ over the different scales. One should note that the large events, or respectively our rogue wave events, are well within this stochastic description.  

\section{Conclusions and Outlook}

We have shown that for ocean wave data measured at a single point a stochastic description can be achieved. It turns out that by means of the empirical mode decomposition and by eliminating the influence of the first intrinsic mode function, the remaining signal obeys Markov properties in its scale evolution. Remarkably, but in agreement with theoretical knowledge on rogue waves derived from nonlinear analysis, the filtered data still contains all large wave, or rogue wave events. The Markov properties in the scale variable $\tau$ indicate that even large scale coherent or correlated structures in the data can be expressed by three point statistics. This is a drastic reduction in complexity. Here one has to keep in mind that the Markov properties were found in the scale variable and not in time. Thus even structures in the signal that correspond to special combinations of several height increments on different scales are grasped by this three point closure, or expressing this stochastically, the joint probability density function $p(h_n, \tau_n;. . .;h_1, \tau_1;h_0, \tau_0)$ can be expressed by the knowledge $p(h_i, \tau_n |  h_{i-1}, \tau_{i-1})$.

A further consequence of the Markov properties is that the scale evolution of the height increments statistics can be expressed by a Fokker-Planck equation, which finally allows to set the complexity of wave structures in the context of non-equilibrium thermodynamics \cite{Nick2013}. For the Fokker-Planck equation drift and diffusion coefficients could be estimated from the data. Based on this a verification of this description shows an excellent agreement between the empirical statistics and the one obtained from the Fokker-Planck equation. Most interestingly also the extreme wave structures seem to be grasped quite well by this approach. This might be a very far reaching finding for a number of reasons. First, the result suggests that in the end also the rogue wave statistics can be captured by relatively simple Fokker-Planck equations, although still with a view to an underlying multi-scale approach involved. Second, in extreme events it is an ongoing discussion, if the very extreme outliers in the data could in any way still be integrated into the fundamental statistics of the complex wave process state. In our present study it seems tempting to conjecture that we might indeed have succeeded in finding such a general approach unifying the rogue and non-rogue waves. This would open also the possibility of a stochastic forecasting of extreme wave events like it was achieved for financial data \cite{Nawroth2010}.

However, although the present findings seem promising, a number of questions definitely require further study. First of all, the present analysis is based on a single data set only, and more data based evidence is without a doubt necessary to gain more and further confidence in the present approach. Second, from a theoretical perspective and a view to the discussion if breather waves or modulation instability and nonlinear focussing could be the backbone structures of rogue waves, more and additional understanding is definitely needed. It seems plausible that the EMD based filtering approach of the data employed in the present study focuses the data analysis towards processes that manifest some local non-trivial phase dynamics and local phase changes, as it is also manifested by breather solutions, modulational dynamics and nonlinear focussing in general. So the present filtering approach might indeed be regarded as a nonlinear transformation of the original data that in the end allows a unified statistical description. Further understanding on why and how this surprising result has been achieved, and if the same result will apply also in other wave states, will have to be gathered. 

In consequence, further work applying the present single point analysis to more wave data including rogue waves has to be conducted. Moreover, also multi-point data will have to be used to capture possibly underlying effects of multi-directional wave states and their role in the present context. From a conceptual perspective, especially the role of the EMD based filtering will have to be clarified to better understand the complex interplay between the linear and nonlinear wave dynamics of the sea state with rogue and non-rogue waves, and a stochastic framework capturing it.       
\ack {The authors wish to express their gratitude to M.R.R. Tabar, P. Milan and A. Chabchoub for helpful discussions. This work was supported by VolkswagenStiftung.}

\end{document}